\documentclass[prb,aps,showpacs,twocolumn]{revtex4}

\usepackage{times}
\usepackage{graphicx}
\usepackage{float}
\usepackage{latexsym,amsmath,amssymb,bm,euscript}
\usepackage{color}
\usepackage{subfigure}
\usepackage{epstopdf}
\usepackage[colorlinks=true,linkcolor=blue,citecolor=blue]{hyperref}

\begin{document}

\title{Possible non-abelian Moore-Read state in double-layer bosonic fractional quantum Hall system}

\author{W. Zhu$^1$, S. S. Gong$^1$,  D. N. Sheng$^{1}$, L Sheng$^2$}
\affiliation{$^1$Department of Physics and Astronomy, California State University, Northridge, California 91330, USA\\
$^2$ National Laboratory of Solid State Microstructures, Department of Physics,
and Collaborative Innovation Center of Advanced Microstructures, Nanjing University, Nanjing 210093, China}

\begin{abstract}
Identifying and understanding interacting systems that can host non-Abelian topological
phases with fractionalized quasi-particles have  attracted intense attention in the past twenty years.
Theoretically, it is possible to realize a rich variety of such states
by coupling two  Abelian fractional quantum Hall  (FQH)  states together through gapping out
part of the low energy degrees of freedom.
So far, there are some indications, but no robust examples have been established in bilayer systems for realizing the
non-Abelian state in the past.
Here, we present a phase diagram of a double-layer bosonic FQH  system based on the
exact diagonalization and density-matrix renormalization group (DMRG) calculations,
which demonstrates a potential regime with the emergence of the non-Abelian bosonic Moore-Read state.
%We start from the Abelian phase with four fold topological degeneracies on torus geometry
%when the two layers are weakly coupled.
%With the increase of interlayer tunneling, we find  an intermediate regime
%with a three-fold groundstate degeneracy and a finite drag Hall conductance.
Starting from the Abelian phase with four fold topological degeneracy at weak coupling,
with the increase of interlayer tunneling, we find an intermediate regime with a three-fold groundstate degeneracy and a finite drag Hall conductance.
We find   different topological sectors in consistent with Moore-Read state
by inserting different fluxes in adiabatic DMRG study.
We also extract the modular $\mathcal{S}-$ matrix, which supports the emergence of the non-Abelian Ising anyon quasiparticle
in this system.
\end{abstract}


\maketitle

\section{Introduction}

The topological states of matter with emergent fractionalized  quasiparticles
have attracted the intense attentions in the past two decades \cite{Wen1990,Stern}.
The  statistics of fractionalized quasiparticles fall into two broad categories:
Abelian \cite{Laughlin1983} and non-Abelian \cite{Moore,Greiter,Read}.
Interchanging two Abelian quasiparticles changes  the groundstates by a non-trivial phase factor,
whereas interchanging two non-Abelian quasiparticles
rotates the groundstates within a set of degenerated groundstate manifold
and the final state will depend on the order of operations being carried out.
Since the quasipartices have unique characterization to a given topological order,
it is possible to identify a topological ordered state through the properties of quasiparticles.
In particular, compared to the Abelian quasipartcles,
the non-Abelian quasiparticles are  fundamentally important for
our understanding of the emerging  physics, which
also have  potential application
for the fault-tolerant topological quantum computation \cite{Kitaev2003,Nayak2008}.
Currently, the most promising platform to investigate the fractional  statistics
is the fractional quantum Hall (FQH) systems, where
most of the observed  FQH states  carry Abelian quasiparticles.
The prominent candidates which may host non-Abelian quasiparticles
include the Moore-Read (MR) Pfaffian state \cite{Moore} at the filling factor $\nu=5/2$  \cite{Willett}
and Read-Rezayi (RR) state \cite{Read} at $\nu=12/5$ \cite{Xia},
which are the
FQH states for electron systems subject to strong magnetic field.

FQH states are also observed in the double-layer systems \cite{Jim1992,Suen1992},
which can be  described in terms of the two-component Halperin states \cite{Halperin1983,Halperin1994}.
Interestingly, the non-Abelian FQH states may also be realized
in the double-layer FQH systems through tuning interlayer
tunneling  and interactions \cite{Ho1995,Nayak1996,Fradkin1999, Wen1999, Read1999, Wen2000,Rezayi2010}.
In such double-layer systems, the Halperin wavefunction upon symmetrization over the layer index
indeed shows the  characteristic features predicted for the non-Abelian states,
including the counting of edge excitations \cite{Rezayi2010} and
the quasihole states \cite{Nayak1996}.
In the theoretical considerations, the non-Abelian state may be induced
by increasing the interlayer coupling, which can gap out the low energy degrees of freedom
that are antisymmetric about the layer inversion
\cite{Fradkin1999, Wen1999, Read1999, Wen2000, Wen2011, Cappelli1999, Cappelli2001, Cabra, Barkeshli}.
However, there are no strong evidence  that this mechanism has  been realized in physical systems.
The numerical studies based on exact diagonalization (ED)
on the  $\nu=1/2$ FQH state for electron systems in double-layer  find a large
wavefunction overlap between the ground state and the non-Abelian state \cite{Regnault2008,Read1996,Papic2010,Peterson2010},
which is consistent with the symmetrization mechanism of the Halperin wavefunction.
However, the obtained energy spectrum has redundant low-energy excitations
without a robust energy gap or the groundstate degeneracy on torus geometry \cite{Papic2010},
which are not consistent with  a non-Abelian QHE state.
Very recently, the double-layer bosonic systems on a lattice model with topological flat bands have been
studied variationally based on the parton construction and Gutzwiller projected wavefunction.
The non-Abelian MR state has been identified
by using the topological spin and chiral central charge \cite{YZhang2014}.
However, it remains an open question whether the non-Abelian state is indeed
the ground state of the microscopic Hamiltonian.
As the possible MR state in this double-layer system appears to be very weak,
the accurate simulations for large systems  and the systematic numerical
characterization of the topological features are highly desired to
pin down the nature of the intermediate phase in these systems.

In this paper, we study a bosonic double-layer system \cite{YZhang2014}
with each layer in the $\nu=1/2$ Laughlin state in the decoupled limit using the ED and
density-matrix renormalization group (DMRG) calculations.
With tuning the tunneling and interaction, we find that the four-fold degenerate ground
states in the decoupled limit evolve to the three  lowest-energy states, which are
symmetry states with respect to the layer inversion.
These low energy states are separated from the higher energy spectrum by a finite gap in an
intermediate parameter regime.
To identify the nature of the intermediate phase,
we design and perform  different flux insertion simulations\cite{Laughlin1981,Prange},
which can identify total Hall  and drag Hall conductances \cite{DNSheng2003}.
We find that the intermediate phase is characterized by the
quantized charge Hall conductance and non-zero drag Hall conductance,
which is distinguished from the Abelian phase with the quantized
charge Hall conductance and zero drag Hall conductance.
The flux insertion studies indicate that with growing interlayer
tunneling, the system evolves from the two-component Abelian phase
to a one-component phase with the nonzero drag Hall conductance, which is consistent with a non-Abelian
state. Furthermore, we calculate the modular $\mathcal{S}$-matrix
using ED on finite-size clusters and  the obtained $\mathcal{S}$-matrix
fully agrees with  that of the MR state.

The remaining of the paper is organized as following:
In Sec. II, we introduce the double-layer lattice model built from
the topological flat-band (TFB) models.
In Sec. III, we present our phase diagram determined by the energy spectrum,
charge Hall conductance (Chern number) and drag Hall conductance.
In Sec. IV, we  present the details of numerical results for the
quantum phase diagram. We show the evolution of the  energy spectrum with tunneling $t_{\perp}$
obtained from ED calculation.
 The robust groundstate degeneracy is checked   by
 inserting  charge  and spin fluxes in ED.
We further establish the quantum phase diagram based on the newly developed  large scale
adiabatical  DMRG simulations, where we
identify  the quantized Chern number and finite drag Hall conductance
%response of the ground state
by inserting  the fluxes in the possible non-Abelian region.
% based on the
%newly developed adiabatical DMRG algorithm.
Furthermore, the modular matrix is calculated to support the emergence of non-Abelian Ising anyon
in the possible non-Abelian regime.
In Sec. V, we discuss the topological trivial phase in our phase diagram. Finally,
in Sec. VI,  we summarize  our main results and
discuss open questions.

\section{Theoretical model}

We consider a double-layer system composed from two single layer
topological flat band (TFB) models
\cite{Haldane1988,Kapit2010,Tang,Neupert2011a,Sun,DNSheng2011,Bernevig2011, YFWang2012}, %on the square lattice,
which can be generally written as:
\begin{eqnarray} \label{Hamiltonian}
\mathcal{H}&=&\mathcal{H}_{\uparrow}+\mathcal{H}_{\downarrow}+\mathcal{H}_{t}+\mathcal{H}_{U},\\
\mathcal{H}_{\uparrow (\downarrow)}&=&-\sum_{(\mathbf{r}\mathbf{r}^{\prime})}
\left[J_{\mathbf{r}\mathbf{r}^{\prime}}e^{i\phi_{\mathbf{r}^{ \prime}\mathbf{r}}}
b^{\dagger}_{\mathbf{r}^{ \prime} \uparrow(\downarrow)}b_{\mathbf{r} \uparrow(\downarrow)}
+\textrm{H.c.}\right], \nonumber\\
\mathcal{H}_{t}&=& t_{\perp}\sum_{\mathbf{r} } \left[ b^{\dagger}_{\mathbf{r} \downarrow} b_{\mathbf{r} \uparrow} + \textrm{H.c.} \right], \nonumber\\
\mathcal{H}_{U}&=& U_{\perp}\sum_{\mathbf{r}} n_{\mathbf{r} \downarrow} n_{\mathbf{r} \uparrow}, \nonumber
\end{eqnarray}
where
$\mathcal{H}_{\uparrow (\downarrow)}$ denotes the hopping terms in the top layer (bottom layer),
$\mathcal{H}_{t}$ describes the interlayer tunneling and
$\mathcal{H}_U$ is the interlayer interaction.
$b^{\dagger}_{\mathbf{r},s} (b_{\mathbf{r},s})$ ($s=\uparrow$ or $\downarrow$)
creates (annihilates) a hard-core boson at site $\mathbf{r}$.
We consider the TFB model on the square lattice and select the phase factor $\phi_{\mathbf{r}^{ \prime}\mathbf{r}}$
corresponding to half flux quanta per plaquette \cite{Kapit2010,Neupert2011a}.
The intralayer hopping terms in $\mathcal{H}_{\uparrow (\downarrow)}$ include
the nearest-neighbor coupling $J_{\langle rr'\rangle}=1.0$ (energy scale), the
next-nearest-neighbor coupling $J_{\langle\langle rr'\rangle \rangle}=0.2941$,
and the next-next-nearest-neighbor coupling $J_{\langle \langle \langle rr'\rangle \rangle \rangle}=-0.2061$,
which give a TFB in each layer with the flatness ratio around $28$.
Such a model can realize the  FQH effect for hardcore bosons or interacting fermions
without a magnetic field due to the  nontrivial Chern number ($C=1$) carried by the topological band
and the reduced kinetic energy \cite{Tang,Neupert2011a,Sun,DNSheng2011,Bernevig2011}.

We consider a finite size system with $2\times N_x\times N_y$ sites ($N_x\times N_y$ sites for each layer)
and the total filling factor of the lower TFB
$\nu=N_p/N_s=(N^{\uparrow}_p+N^{\downarrow}_p)/N_s=1$ \cite{Cooper2001,Regnault2003},
where $N_p$ is the total number of hardcore bosons and $N_{s}$ is the number of single-particle states of the lower TFB.
IN the following, we will refer the layer index as the (pseudo)spin index for convenience.
In the absence of tunneling, the system reduces to the decoupled two layers with each
layer at the $1/2$ filling.
While the ED calculations on torus geometry are limited to the  system with  $40$ sites,
DMRG \cite{SWhite1992} allows us to study the systems up to $256$ sites on the cylinder
 which have the periodic boundary in one direction and the open boundary
in the other direction.
We keep up to $M = 3600$ states in DMRG calculations, which give accurate results.

\begin{figure}[t]
 \begin{minipage}{0.95\linewidth}
 \centering
 \includegraphics[width=2.5in]{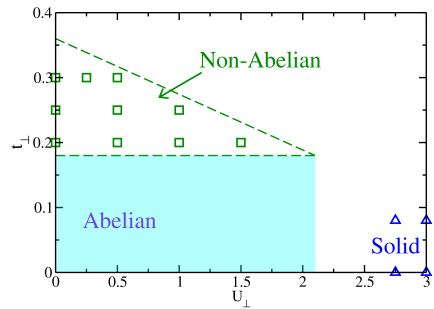}
 \end{minipage}
\caption{(Color online)
Quantum phase diagram of the double-layer system for the Hamiltonian
Eq. (\ref{Hamiltonian}) with changing  the interlayer tunneling $t_{\perp}$ and interaction $U_{\perp}$.
The blue color filled region represents the topological Abelian phase with the
fourfold degenerate ground states and the topological Chern Number $C^{c} = 1$.
The green colored squares indicate an intermediate topological order phase with
threefold degeneracy and $C^{c} = 1$,
which is identified as the one-component non-Abelian Moore-Read state by the finite drag Hall
conductance and the modular $S$-matrix. The yellow dots represent a solid phase that breaks lattice symmetries.
}\label{fig:phase}
\end{figure}

\section{Phase diagram}
Our main results are summarized as the phase diagram in the
$t_{\perp}-U_{\perp}$ plane  as shown in Fig. \ref{fig:phase}.
We find that the Abelian FQH state of single layer ($t_\perp = U_\perp = 0$) is stable against
the weak interlayer tunneling. In the double-layer system, this Abelian phase
is characterized by the robust fourfold degenerate ground states,
the quantized total charge Chern number $C^c=1$, and the vanishing-small drag Hall conductance.
With the increase of interlayer tunneling on finite-size system, the energy of
the single antisymmetric  ground state splits from  the other three
symmetrized ground states. The higher energy spectrum has a gap
from the threefold symmetrized ground states
in an intermediate $t_\perp$ region (for $U_\perp = 0$, $0.2 \lesssim t_\perp \lesssim 0.3$).
In both the Abelian phase and the intermediate region, we find that
the charge Chern Number is always quantized as $C^{c} = 1$; however,
the drag Hall conductance jumps from the vanishing-small value in the Abelian
phase to the almost saturated value in the intermediate region, indicating
that the system evolves from a two-component to a one-component QHE  state.
The phase boundary between Abelian and the possible non-Abelian phase is
determined by drag Hall conductance reaching half of its saturated value ($\sim 0.125$)
based on DMRG simulation.
Interestingly, using the threefold  ground states, we
extract the modular $\mathcal{S}-$matrix, which supports
a non-Abelian MR state with the emergent Ising anyon quasiparticle and the corresponding fusion rule.
In the larger tunneling regime ($t_{\perp} > 0.3$), the system becomes a
topologically trivial state with vanishing Chern numbers ($C^c=0$).
In the large $U_\perp$ region, we find a charge density wave state that
breaks lattice symmetries.

\begin{figure}[t]
 \begin{minipage}{0.9\linewidth}
 \centering
 \includegraphics[width=3.0in]{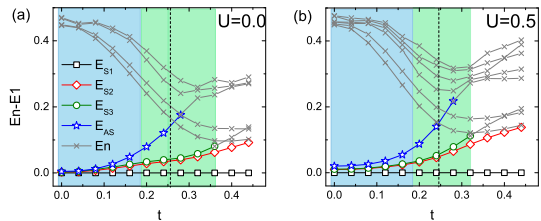}
 \end{minipage}
 \caption{(Color online) Energy spectrum evolution with the interlayer tunneling $t_{\perp}$ on $2\times 4\times 4$ torus geometry using ED calculation for (a) $U_{\perp}=0.0$ and (b) $U_{\perp}=0.5$.
Among  the four near degenerating ground states,
the three symmetric (S1,S2,S3) states are  labeled by the
black, green, and red circles, while the  anti-symmetric (AS) state is labeled by the blue stars.
}\label{fig:ED:energy}
\end{figure}

\section{Topological nontrivial quantum Hall  phases}
\subsection{Energy spectrum on torus}
We first use ED to study the evolution of energy spectrum with tunneling $t_{\perp}$ at $U_{\perp}=0$
on the $2\times 4\times 4$ torus system
At $t_{\perp}=0$, both  layers have
the bosonic $\nu=1/2$ Laughlin states with twofold degenerate ground states.
Thus, the energy spectrum of the whole  system has a fourfold degeneracy
separated from higher energy levels by a robust gap.
As the double layer system has a layer inversion symmetry \cite{Read1999,Papic2010,Peterson2010,Nomura},
the fourfold ground states can be  divided into two groups:
the symmetric group with three states $E_{S1},E_{S2},E_{S3}$ shown as open circles
and the anti-symmetric group with a single state $E_{AS}$ represented by  blue stars
as shown in Fig. \ref{fig:ED:energy}.
By increasing the interlayer tunneling, the groundstate degeneracy is
lifted gradually. The energy of the anti-symmetric state  grows rapidly and merges into the high energy
continuum at $t_{\perp}\approx 0.25$.
With further increasing $t_{\perp}$, two of the three symmetric states also merge into high energy continuum
at $t_{\perp}\approx 0.36$.
In the intermediate regime $0.20\lesssim t_{\perp}\lesssim 0.36$,
the three symmetric states have  close energies separating from
 the higher energy anti-symmetric state by a finite gap.
Although there is a finite splitting between
the lowest three states due to the finite size effect allowing
topological distinct states being coupled together, the low-energy spectrum of the intermediate region
implies a possible threefold degenerate ground states protected by a gap in the thermodynamic limit.
In the presence of interlayer interaction,
we observe the similar results
as shown in Fig.~\ref{fig:ED:energy}(b) for $U_{\perp}=0.5$.
Thus, the phase region with $t_{\perp}\approx 0.25$, which has the maximum finite-size
gap, may  be the suitable phase regime for observing  the possible non-Abelian Moore-Read state.

\begin{figure}[b]
 \begin{minipage}{0.9\linewidth}
 \centering
 \includegraphics[width=3.0in]{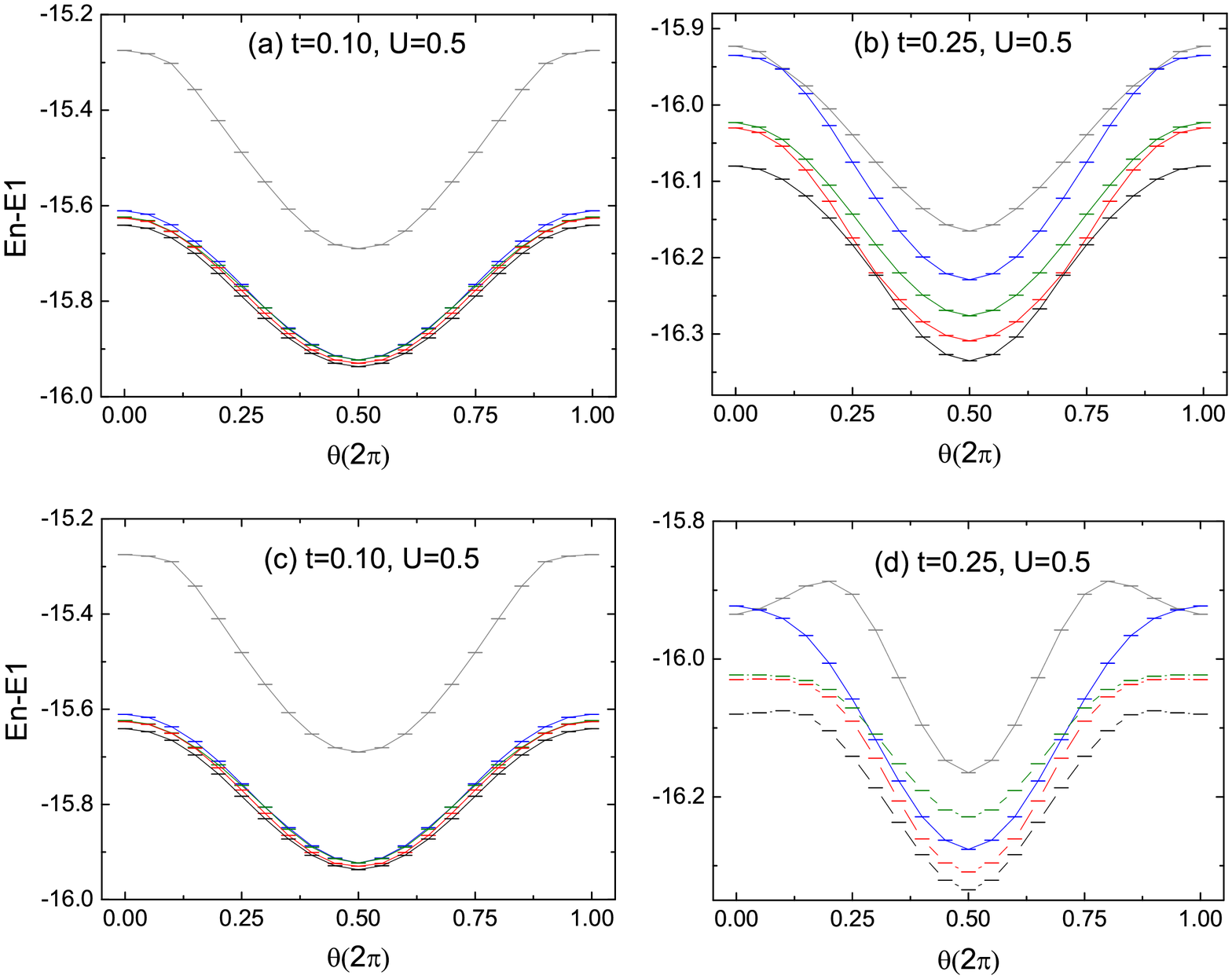}
 \end{minipage}
 \caption{(Color online) Energy spectrum evolution with (a-b) inserting a charge flux $\theta^{\uparrow}_y=\theta^{\downarrow}_y$ and
 (c-d) a spin flux $\theta^{\uparrow}_y=-\theta^{\downarrow}_y$ for $(t_{\perp},U_{\perp})=(0.1,0.5)$ and
 $(t_{\perp},U_{\perp})=(0.25,0.5)$. The lowest five energy levels are shown.
}\label{fig:ED:flux}
\end{figure}

\subsection{Flux insertion based on ED}
We  check  whether the threefold degeneracy is robust
by considering the response of the low-energy
spectrum to the flux insertion. To induce the flux,
we impose a twisted boundary condition in the $\hat{y}$-direction:
$\langle\mathbf{r_i}+N_y\hat{y}|\Psi_{\theta_y}\rangle=\langle\mathbf{r_i}|e^{i\theta_y \sigma_{0(3)}}|\Psi_{\theta_y}\rangle$,
where $\Psi_{\theta_y}$ is the many-body state with boundary phase $\theta_y$
and Pauli matrix $\sigma_{0(3)}$ acts on the layer degrees of freedom of the particle
at the position $r_i$.
The twisted boundary is equivalent to threading a flux in the hole of a torus along the $\hat{x}$-direction \cite{QNiu}.
In the double-layer system, we introduce two kinds of boundary conditions:
charge flux ($\sigma_0$) and spin flux ($\sigma_3$) \cite{DNSheng2003, DNSheng2006, Neupert2011b}.
In the charge and spin flux,
the boundary phases in the top and bottom layers have the same ($\theta^{\uparrow}_y=\theta^{\downarrow}_y$)
and the opposite  ($\theta^{\uparrow}_y=-\theta^{\downarrow}_y$) signs, respectively.
For the topological states, the degenerate ground states should remain gapped without
 level crossing with the higher energy levels in the charge flux insertion.
Therefore, the charge flux insertion can be used to identify the near degenerate ground states
from the low-energy levels.
Furthermore, it is expected that a two-component double-layer system
will have  similar responses to the charge and spin fluxes, while
a coupled one-component system has the different responses.

Fig. \ref{fig:ED:flux} shows the ED results of the evolution
of low-energy spectra with inserting charge and spin fluxes for
the system with weak and intermediate tunnelings.
For $t_{\perp}=0.10$ and $U_{\perp}=0.5$,
the four lowest-energy states are always protected by a gap
with tuning the charge (Fig. \ref{fig:ED:flux}(a)) or the spin flux (Fig. \ref{fig:ED:flux}(c)),
which are consistent with a two-component topological state with fourfold degeneracy.
In the possible non-Abelian phase for $(t_{\perp},U_{\perp})=(0.25,0.5)$,
the lowest three energy levels are always separated from the higher energy spectrum by a small gap
in the charge flux insertion (Fig. \ref{fig:ED:flux}(b)).
By inserting a spin flux (Fig. \ref{fig:ED:flux}(d)),
we find that the lowest three levels connect with higher energy states.
These observations indicate that the intermediate region is a strongly coupled one-component
topological state with threefold degeneracy.
However, the nature of the intermediate regime can
not be established based on these ED calculations.
As shown in the phase diagram Fig.~\ref{fig:phase}, we determine the  phase boundary of the
Abelian phase where the fourfold degeneracy of energy spectrum is being destructed,
and  a possible non-Abelian phase may emerge.
We will present the full evidence of the nature of the non-Abelian state below.

\begin{figure}[t]
 \begin{minipage}{1.0\linewidth}
 \centering
 \includegraphics[width=3.5in]{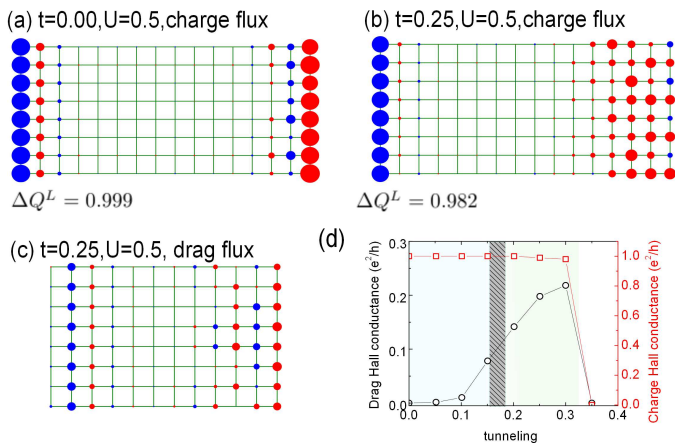}
 \end{minipage}
 \caption{(Color online) Real-space configuration of charge accumulation $\langle n^{\uparrow}_i+n^{\downarrow}_i\rangle$
 after adiabatically inserting a unit charge flux  $\theta^{\uparrow}_y=\theta^{\downarrow}_y=2\pi$ for
  (a) $(t_{\perp},U_{\perp})=(0.00,0.5)$ and (b) $(0.25,0.5)$.
 The area of the circle is proportional to the accumulation amplitude and the red (blue) color represents the positive (negative) value.
  The results are calculated on $2\times 8\times 16$ cylinder using DMRG by keeping $3600$ states.
  (c) Real-space configuration of charge accumulation $\langle n^{\downarrow}_i\rangle$
   after adiabatically inserting a quantized drag flux  $\theta^{\uparrow}_y=2\pi  (\theta^{\downarrow}_y=0)$.
 (d) The Charge Hall conductance (red squares) and drag Hall conductance (black circles) versus tunneling $t_{\perp}$.
 The results for drag flux are calculated on $2\times 8\times 12$ cylinder using DMRG by keeping $3600$ states.
}\label{fig:dmrg:chern}
\end{figure}

\subsection{Flux insertion based on DMRG}
To further investigate the topological properties of the double-layer
system on larger size, we use DMRG to study the cylinder
system by adiabatically threading a charge or drag flux.
The drag flux is realized by introducing the twist boundary phase in just one layer.
which can induce a Hall response in its  own layer and also drag the particles in the other layer.
Theoretically, the drag Hall conductance  and its connection to the topological Chern number matrix
\cite{QNiu,DNSheng2003,DNSheng2006} has been established before.
Conventionally, one obtains such topological Chern invariants based on ED calculations \cite{DNSheng2003,DNSheng2006}. Very recently, the flux insertion  has been introduced in
 the large-scale DMRG  simulation on cylinder systems\cite{He2013,SSGong2013,Zaletel},
which can be used to detect different Hall conductances.

Very interestingly, the inserting fluxes method
we establish here for double layer systems
can also be used to access all the topological sectors in the  system.
%%% we find that by applying both the charge and drag fluxes in DMRG calculations,
This argument can be easily understood in the decoupled limit.
Starting from the ground state without any flux, we insert the charge
flux by adiabatically increasing the twist boundary
phase in the closed boundary along the $y$ axis
$\theta^{\uparrow}_y = \theta^{\downarrow}_y = 0 \rightarrow 2\pi$.
By inserting  $2\pi$ flux, the ground state of each layer
evolves into a new topological sector with a fractional 1/2 charged  quasi-particle being
pumped from one edge of cylinder to the other one \cite{SSGong2013}.
Then, by adding the drag flux in either layers separately, the system would evolve
to the other two sectors, which has one more pumped charge $1/2$ quasi-particle
in the layer with drag flux. Therefore, we obtain the four topological degenerate
ground states in the Abelian phase.
Qualitatively, this picture in the decoupled limit applies to the whole Abelian phase
as long as the drag Hall conductance is vanishing small, i.e., one layer cannot effectively
drag the other layer. When the system has a transition from two components to one component \cite{KunYang},
the drag Hall conductance would jump to a finite value close to the saturated value.
In the coupled one-component system,  the drag flux
applied to either one of the  layer will evolve the system  to the same topological sector by pumping
one fractional charged quasiparticle,  thus we only obtain three topological sectors for coupled phase.

As shown in Fig. \ref{fig:dmrg:chern}(a), after the insertion of one charge flux quantum
($\theta^{\uparrow}_y=\theta^{\downarrow}_y=0\rightarrow 2\pi$),
a net quantized charge (boson number) accumulates at
the left edge ($n_c^{\uparrow}+n_c^{\downarrow}\approx 0.999$ is the net charge accumulation).
The charge accumulation is equivalent to a net charge transfer from the right edge to the left
edge. According to the fundamental correspondence between edge transfer and bulk Chern number \cite{Prange},
we find a quantized charge Chern number $C^{c}=1$ for this Abelian FQH state.

\begin{figure}[t]
 \begin{minipage}{0.9\linewidth}
 \centering
 \includegraphics[width=3.5in]{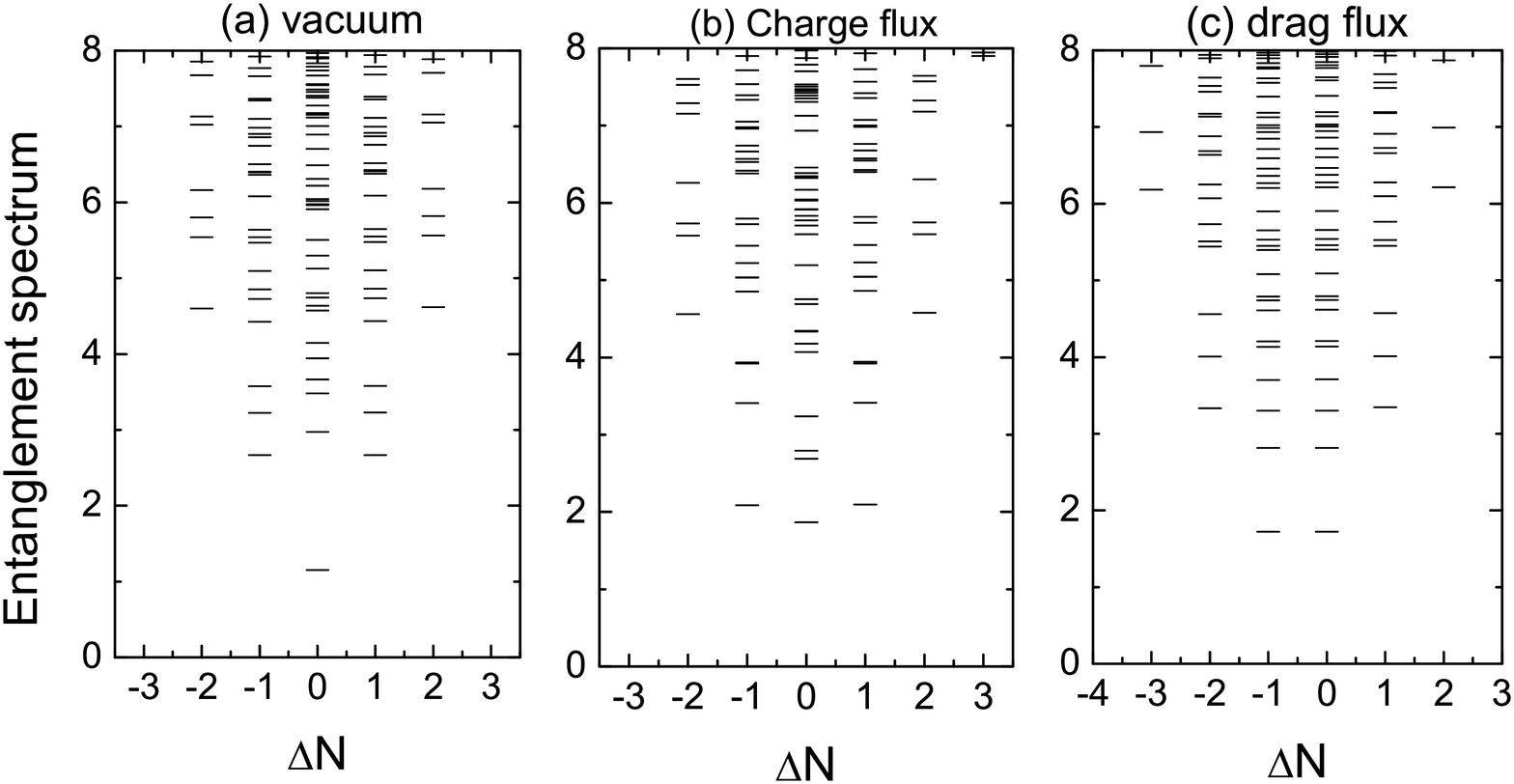}
 \end{minipage}
 \caption{(Color online)  The entanglement spectrum for $t_\perp = 0.25, U_\perp = 0.0$ on
  the $L_y=8$ cylinder with (a) no flux, (b) a charge flux quanta, and (c) a drag flux (flux imposed on one layer only).
 The entanglement spectrum are grouped by the relative boson number $\Delta N$ of left half cylinder.
 The results are obtained using DMRG by keeping up to $1600$ states.
}\label{fig:dmrg:ES}
\end{figure}

The quantized charge transfer also persists in the possible non-Abelian parameter region.
As shown in Fig. \ref{fig:dmrg:chern}(b),
the net charge transfer corresponds to a quantized charge Chern number $C^{c}=1$,
which indicates the topological nontrivial nature of the possible non-Abelian phase.
Nevertheless, the charge flux cannot distinguish the Abelian phase from the possible non-Abelian phase since both of them share the same  $C^{c}=1$.
To further identify the phase transition between the Abelian phase and the possible non-Abelian phase,
we consider the effects of the drag flux.

By threading a drag flux quantum in the top layer
($\theta^{\uparrow}_y=0 \rightarrow 2\pi$, $\theta^{\downarrow}_y=0$),
we observe the  particle accumulations in the bottom layer \cite{DNSheng2003,KunYang}
as shown in Fig.~\ref{fig:dmrg:chern}(c)). By calculating the
drag Hall conductance as a function of  $t_{\perp}$ as displayed in
Fig. \ref{fig:dmrg:chern}(d),
we find a strong enhancement of drag Hall conductance at $t_\perp \simeq 0.15$,  which coincides with the phase
boundary of the disappearance of the Abelian phase identified  from the  ED energy spectrum calculations.
Within the regime $0.25 \lesssim t_{\perp} \lesssim 0.30$, the drag Hall conductance approaches
the saturated value $0.25$, which is consistent with an effective   one-component system.
Based on the above results, we determine the Abelian phase (blue color), the
one-component  non-Abelian phase (green squares)
as shown  in Fig. \ref{fig:phase}.
The topological trivial solid phase with  zero charge Chern number is also shown.

To further identify the possible non-Abelian state, we investigate the
entanglement spectra of each topological sector. Physically, the different
topological ground states on a cylinder are expected to have the different
well-defined anyonic flux through the cylinder. Thus, the cylinder system
with the charge flux or the drag flux corresponds to the other two topologically distinct ground states
besides the vacuum state.
To explicitly demonstrate the ground states with different anyonic flux on cylinder geometry,
we bipartite the cylinder into two halves, and observe entanglement spectrum \cite{Haldane2008}
to distinguish the different topological sectors.
As shown in Fig. \ref{fig:dmrg:ES}, we show the entanglement spectrum for
the vacuum ground state in Fig. \ref{fig:dmrg:ES}(a), the new ground state obtained
by inserting a charge flux quantum in Fig. \ref{fig:dmrg:ES}(b), and the
ground state obtained by inserting a drag flux in Fig. \ref{fig:dmrg:ES}(c).
These three ground states are anticipated to have one-to-one correspondence
with identity, fermion, and Ising anyon sectors, respectively.
We also calculate the momentum dependence of the entanglement spectra in each $U(1)$
quantum number sector with different relative boson number $\Delta N$, and obtain
the counting of the leading eigenvalues in the entanglement spectra \cite{Vidal}.
The obtained results are
similar to those of coupled two Laughlin $\nu=1/2$ states.
Due to the calculation limit, $N_x=8$ (16 lattice sites in the $\hat{x}$ direction)
is the largest width we can reach convergence in our  DMRG calculations,
which gives four momentum quantum numbers  $K=0,\pi/2,\pi,3\pi/2$ in each $\Delta N$ sector.
Although we observe that a very small entanglement gap opens up in $K=\pi$ and $3\pi/2$ sectors
between the expected counting for non-Abelian MR state
and the other part of entanglement spectrum,
we cannot determine if the gap will  survive in the thermodynamic limit or it is a finite size effect.
Since all other results support the non-Abelian QHE state,  we believe this result
is due to the
finite size effect and we leave this part for the future study.

\begin{figure}[b]
 \begin{minipage}{0.49\linewidth}
  \includegraphics[width=1.5in]{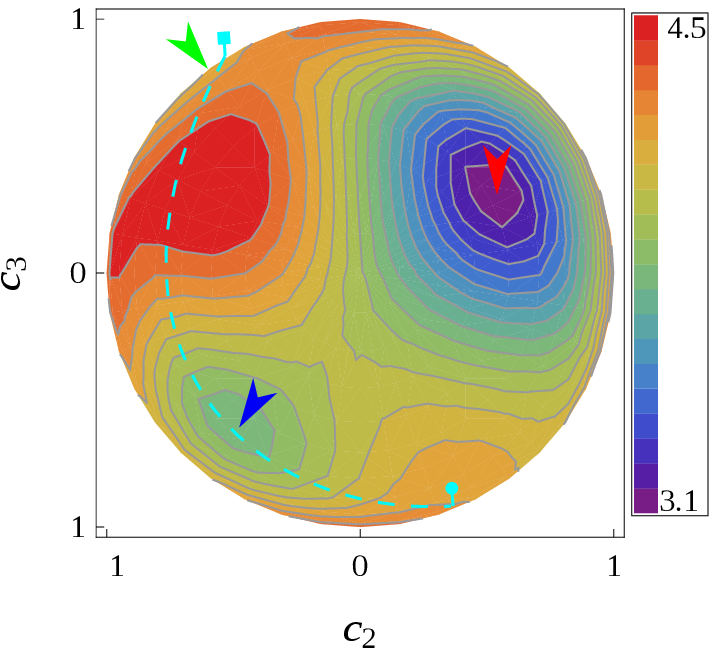}
 \end{minipage}
  \begin{minipage}{0.45\linewidth}
   \includegraphics[width=1.4in]{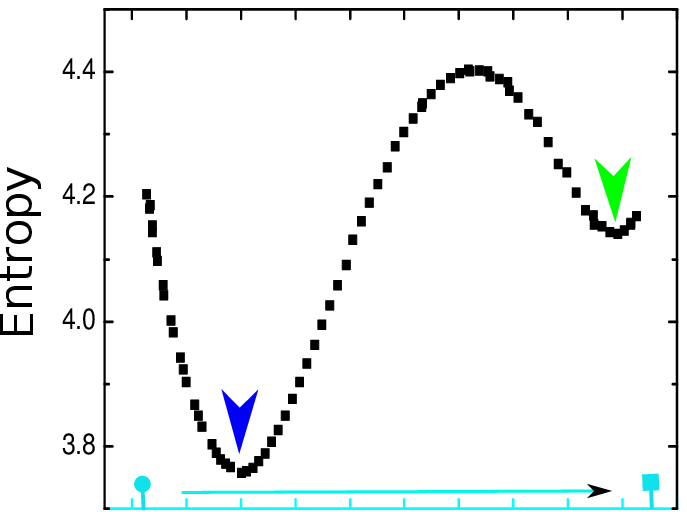}
  \end{minipage}
 \caption{(Color online)
  Left: Contour plot of entanglement entropy of $|\Psi_{c_1,c_2,c_3,\phi_2,\phi_3}\rangle$
  on $2\times 4\times 4$ system.
  We show entropy profile versus $c_2,c_3$ ($c_1=\sqrt{1-c_2^2-c_3^2}$)
  by setting optimized $\phi^o_2,\phi^o_3$.
  Three nearly orthogonal MESs are marked by red, green and blue arrows in surface plot.
  The cyan dashed line represents the states orthogonal to the first MES (red arrow).
  Right: Entropy for the states along the cyan dashed line as shown in left figure.
}\label{fig:ED:MES}
\end{figure}

\subsection{Modular matrix}
From the above observation of  DMRG, we find that the intermediate phase region
appears as a one-component topological nontrivial phase. Here
we calculate the modular $\mathcal{S}-$matrix using the near degenerate
threefold states in the ED energy spectrum to further investigate the nature
of the possible non-Abelian phase \cite{Wen1990}.
Modular $\mathcal{S}-$matrix encodes the information of the quasiparticle statistics
including quantum dimension and fusion rules \cite{Verlinde,ZHWang,Fendley},
which has been successfully used to identify various Abelian and non-Abelian topological orders \cite{SDong,YZhang2012,YZhang2013,Vidal,WZhu2013,Pollmann,WZhu2014}.
To calculate the modular matrix, we follow the method based on the minimal entangled states (MESs) \cite{YZhang2012}.
The MESs are the eigenstates of the Wilson loop operators
with a definite type of quasiparticle \cite{SDong}. Thus, the modular transformations on the MESs
give rise to the modular matrix \cite{YZhang2012}.

Here we show the results at $t_{\perp}=0.25,U_{\perp}=0.0$ as an example.
We denote the three lowest-energy states in ED spectrum
as $|\xi_j\rangle$ ($j=1,2,3$),
from which we can form the general superposition states as,
\begin{equation*}
|\Psi_{(c_2,c_3,\phi_2,\phi_3)}\rangle=c_1|\xi_1\rangle+c_2e^{i\phi_2}|\xi_2\rangle+c_3e^{i\phi_3}|\xi_3\rangle
\end{equation*}
where $c_1,c_2,c_3$, $\phi_2$, $\phi_3$ are real superposition parameters.
For each state $|\Psi\rangle$, we construct the reduced density
matrix and obtain the corresponding entanglement entropy.
To find the MESs, we optimize the superposition parameters to find the minimum
entanglement entropy. As shown in Fig.~\ref{fig:ED:MES}(a),
we show the entropy profile of $|\Psi\rangle$ with the optimized parameters $(\phi^o_2,\phi^o_3)$
for the middle cut along the $x$-direction. We find the
first global MES $|\Xi^{I}_1\rangle$ with the entropy $S\sim 3.10$ at the position pointed by the red arrow.
The second MES $|\Xi^{I}_2\rangle$ (blue arrow) and the third MES $|\Xi^{I}_3\rangle$ (green arrow) can be determined
in the state space orthogonal to $|\Xi^{I}_1\rangle$, as shown in Fig.~\ref{fig:ED:MES}(b).
Finally,  we can obtain the modular matrix $\mathcal{S}=\langle\Xi^{II}|\Xi^{I}\rangle$  extracted from the overlap between the MESs
for two noncontractible partition cut directions \cite{YZhang2012}:
\begin{eqnarray*}\label{MR:modularS}
&&\mathcal{S}\approx \mathcal{S}^{CS}+\Delta \mathcal{S} = \\
&&\frac{1}{2}
\left(\begin{array}{ccc}
        1 & 1 & \sqrt{2} \\
        1 & 1 & -\sqrt{2} \\
        \sqrt{2} & -\sqrt{2} & 0
       \end{array}
     \right)+ 10^{-2}\times
\left(\begin{array}{ccc}
        -5.6 & 5.4 & 1.4 \\
        5.4 & -2.5 & 3.2 \\
        1.4 & 3.2 & 6.2
       \end{array}
     \right),
\end{eqnarray*}
where $\mathcal{S}^{CS}$ represents the theoretical prediction from the $SU(2)_2$
Chern-Simons theory \cite{SDong,ZHWang,Fendley}.
$\mathcal{S}^{CS}$ determines the quasiparticle quantum dimension as
$d_{\openone}=1$, $d_{\psi}=1$, $d_{\sigma}=\sqrt{2}$ and non-trivial
fusion rule as $\sigma\times \sigma = \openone + \psi$,
where $\openone$ represents the identity particle,
$\psi$ the fermion-type quasiparticle, $\sigma$ the Ising anyon quasiparticle.
Thus, the numerical extracted modular $\mathcal{S}-$matrix identifies the intermediate
topological phase with threefold ground state degeneracy as the non-Abelian MR
state with the emergence of the Ising anyon quasiparticles satisfying the non-Abelian
fusion rule ($\mathcal{S}_{33}\approx 0$)\cite{YZhang2013,WZhu2014}.

Generally speaking, to uniquely determine a topological order, one
needs both the modular $\mathcal{S}$ and $\mathcal{U}$ matrices \cite{YZhang2012}.
From modular $\mathcal{U}$ matrix, one can access the chiral central charge $c$
and topological spin of each quasiparticle, which  distinguish the
non-Abelian MR state from double-layer Abelian Laughlin states.
For example, the chiral central charge of non-Abelian MR state is $c=3/2$, while
the double-layer Laughlin state has $c=2$.
Unfortunately, in the current lattice model (in Eq. \ref{Hamiltonian} ),
the MES route can not give the  $\mathcal{S}$ and $\mathcal{U}$ matrix together since there is no rotation $\pi/3$ symmetry here \cite{YZhang2012}.
Recently, we note that it has been proposed a general method, named momentum polarization\cite{YZhang2014},
to extract the quasiparticle statistics in modular $\mathcal{U}$ matrix.
In this method, one needs to perform a finite-size scaling on $L_y$,
then the statistics information in modular $\mathcal{U}$ matrix can be extracted from $L_y\rightarrow0$ limit.
Unfortunately, in current DMRG calculation we are limited to the system sizes $L_y=4, 6, 8$ due to the
computational capability.
Thus a reliable application of momentum polarization method here is very challenging, which
we leave  for  the future study.
%It is interesting to further apply
%the method to our model using DMRG on cylinder geometry, which we leave for  the future study.

\begin{figure}[t]
 \begin{minipage}{0.9\linewidth}
 \centering
 \includegraphics[height=1.5in]{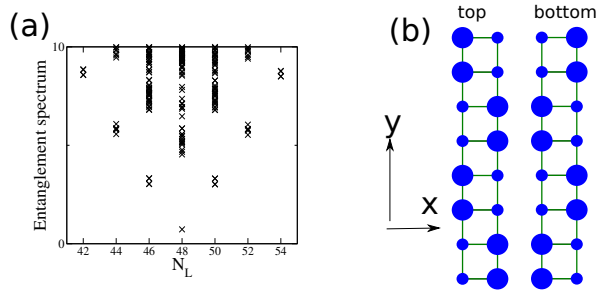}
 \end{minipage}
 \caption{(Color online) (a) Entanglement spectrum for $t_{\perp}=0.08,U_{\perp}=1.2$. (b) The charge density distribution of $\langle n^{\uparrow}_i\rangle$ (top layer) and $\langle n^{\downarrow}_i \rangle$ (bottom layer) in real-space.
 Here we only show two columns in the bulk in each layer. The two-column pattern is periodic along the cylinder direction. The area of the circle is proportional to the amplitude of charge density.
 The results are calculated on the $2\times 8\times 16$ cylinder using DMRG with keeping $1600$ states.
}\label{fig:dmrg:CDW}
\end{figure}

\section{Topological trivial phases}

Besides the two topological non-trivial phases in the phase diagram Fig. \ref{fig:phase},
we also find the topological trivial phases at the large $U_{\perp}$ parameter region. %$t_{\perp}$ or $U_{\perp}$ parameter regions.
Here we show that there is a charge density wave phase at large $U_{\perp}$ region,
as shown by the yellow triangular in Fig. \ref{fig:phase}.
As shown in Fig. \ref{fig:dmrg:CDW} (a),
the entanglement spectrum has a large weight on the lowest eigenvalue, which is a feature of a solid phase.
In Fig. \ref{fig:dmrg:CDW}(b), we demonstrate the charge density wave pattern in real space.
Along the  cylinder axis ($x-$direction), the unit cell (enclosing $4$ sites) is doubled in each layer
due to the charge density pattern.
In the $y-$direction, the charge density pattern has a period  of two.
At each site, the top and bottom layers have the opposite density pattern, which
leaves $\langle n^{\uparrow}_i+n^{\downarrow}_i\rangle$  a constant.

\section{Summary and discussion}

We have studied a bosonic double-layer system on a square lattice
using ED and DMRG calculations. Through the studies of the energy spectrum,
the flux insertion on cylinder, and the modular matrix,
we find numerical evidences for a non-Abelian Moore-Read state
emerging from the bilayer Halperin states through
gapping out the interlayer anti-symmetric state.
Although this practically powerful route to a variety of non-Abelian quantum states
has been introduced theoretically for decades based on parton construction and
field theory \cite{Fradkin1999, Wen1999, Read1999, Wen2000, Wen2011, Cappelli1999, Cappelli2001, Cabra, Barkeshli},
there were limited numerical evidence to support the realization of non-Abelian state
in  microscopic systems.
Our numerical calculations rely on the insertion of charge and drag fluxes,  which
allow us to detect the quantum phase transition from a two-component topological state
to a one-component state characterized by the onset of the finite drag Hall conductance.
In combining with the modular matrix simulation for the quasiparticle statistics,
we identify  the  nature of the intermediate $t_{\perp}$ (with threefold near degeneracy)
phase as the non-Abelian Moore-Read state, although this state is relatively weak and
the entanglement spectrum does not show a robust entanglement spectrum gap for the
counting associated with the non-Abelian state.

%Another important  question is about the nature of the phase transition between the Abelian phase and the possible non-Abelian phase.
We have also explored the nature of the
quantum phase transition  between the Abelian phase and the possible non-Abelian phase.
We have studied quantities such as entanglement entropy and  the wavefunction overlap. We find all of these quantities
of  the groundstate change smoothly when the system crosses the phase boundary. % from Abelian phase to the possible non-Abelian phase.
This indicates that  the phase transition is either weakly first order or continuous transition.
%%%% Another possibility is the intermediate regime is just a crossover from Abelian phase to topological trivial phase and there is no non-Abeiian phase here.
%%%We leave this challenge issue to the future studies.
Moreover, it would be particularly interesting to study the possibility of realizing
the non-Abelian phase from coupled bilayer Halperin states
in fermionic systems,  which will be investigated in the future work.
\\

\textit{Note added.} Upon finalizing the manuscript we noticed several preprints  focusing on
double-layer $\nu=1/3 + 1/3$ fermionic systems\cite{Jeong2014,Geraedts2015,Peterson2015,ZLiu2015}.
%which have identified non-Abelian states in coupled bilayer systems.

\section{Acknowledgements}
WZ thanks Y. Zhang and N. Regnault for insightful comments.
This work is supported by the U.S. Department of Energy,
Office of Basic Energy Sciences under Grant No. DE-FG02- 06ER46305 (WZ, DNS),
the NSF grant DMR-1408560 (SSG),
the State Key Program for Basic Researches of China under grants numbers 2015CB921202, 2014CB921103, and the National Natural Science
Foundation of China under grant numbers 11225420 (LS).

%\clearpage
%\appendixpage

%\begin{appendices}

%\end{appendices}

\end{document}